\documentclass[11pt]{article}
\usepackage[textwidth=15.2cm,textheight=22cm]{geometry}
\usepackage{amsmath,amssymb}
\usepackage{latexsym}
\usepackage{multicol}
\usepackage{graphicx}
\usepackage{caption}
\usepackage{bm}
\numberwithin{equation}{section}
\pdfoutput=1
\usepackage[T1]{fontenc}
\usepackage[latin9]{inputenc}
\usepackage[active]{srcltx}
\usepackage{esint}
\usepackage{cancel}
\usepackage{color}
\usepackage{ulem}
\allowdisplaybreaks[1]

\newcommand{\del}{\partial}
\newcommand{\be}{\begin{equation}}
\newcommand{\ee}{\end{equation}}
\newcommand{\bea}{\begin{eqnarray}}
\newcommand{\eea}{\end{eqnarray}}

\newcommand{\half}{\frac{1}{2}}
\newcommand{\nn}{\nonumber}

\newcommand{\ndt}{\noindent}

\newcommand{\parm}{\partial_-}

\newcommand\fr[1]{\frac{1}{#1}}

\definecolor{ggreen}{rgb}{0.0, 0.5, 0.20}

\begin{document}

	\begin{titlepage}
		\begin{flushright}    
			{\small $\,$}
		\end{flushright}
		\vskip 1cm
		\centerline{\Large{\bf{BMS algebra from residual gauge invariance}}}
		\vskip 0.5cm
		\centerline{\Large{\bf{in light-cone gravity}}}
		 \vskip 1.5cm
		\centerline{Sudarshan Ananth$^\dagger$, Lars Brink$^*$ and Sucheta Majumdar$^\ddagger$}
		\vskip .7cm
		\centerline{$\dagger$\,\it {Indian Institute of Science Education and Research}}
		\centerline{\it {Pune 411008, India}}
		\vskip 0.7cm
		\centerline{$^*\,$\it {Department of  Physics, Chalmers University of Technology}}
		\centerline{\it {S-41296 G\"oteborg, Sweden}}
		\vskip 0.7cm           
		\centerline{$\ddagger$\,\it{Universit\'e Libre de Bruxelles and International Solvay Institutes,}}
		\centerline{\it{Campus Plaine CP231, B-1050 Brussels, Belgium}}
		\vskip 1.5cm
		\centerline{\bf {Abstract}}
		\vskip .5cm
				
		\ndt We analyze the residual gauge freedom in gravity, in four dimensions, in the light-cone gauge, in a formulation where unphysical fields are integrated out. By checking the invariance of the light-cone Hamiltonian, we obtain a set of residual gauge transformations, which satisfy the BMS algebra realized on the two physical fields in the theory. Hence, the BMS algebra appears as a consequence of residual gauge invariance in the bulk and not just at the asymptotic boundary. We highlight the key features of the light-cone BMS algebra and discuss its connection with the quadratic form structure of the Hamiltonian.
		\vfill
	\end{titlepage}
	
	\tableofcontents
	
	\section{Introduction}
	
	\ndt	
	The Bondi-van der Burg-Metzner-Sachs (BMS) group is an infinite-dimensional enhancement of the Poincar\'e group, that arises as the asymptotic symmetry group at null infinity for asymptotically flat spacetimes~\cite{Bondi:1962px,Sachs:1962zza}. In~\cite{Barnich:2009se}, the BMS group was extended to include superrotations. These asymptotic symmetries have been related to soft theorems for gauge theories in the recent years~\cite{Strominger:2017zoo}, following which there has been a renewed interest in the study of these symmetries.
	
\vskip 0.2cm
	\ndt	
	The study of asymptotic symmetries is very sensitive to the boundary conditions and gauge choices imposed on the fields. At spatial infinity, for instance, only the Poincar\'e algebra (and not the BMS algebra) can be canonically realized under the standard boundary conditions in the Hamiltonian formulation of general relativity~\cite{Regge:1974zd}. Recent work however showed that, by relaxing those boundary conditions, the BMS group can be recovered at spatial infinity,  thereby, resolving the puzzle between the asymptotic structure at spatial and null infinity~\cite{Henneaux:2018cst,Henneaux:2019yax}. This raises an important question: How much of the gauge freedom can be fixed in the theory without losing the residual reparameterizations that are associated with the BMS symmetry?
	
	\vskip 0.2cm
	\ndt
	In the light-cone formulation of gravity in four dimensions, one can gauge away the unphysical degrees of freedom and describe the dynamics of the theory in terms of the two physical states of the graviton. In the usual Bondi gauge, the BMS group appears as the asymptotic symmetry of gravity at infinity. In light-cone gravity, we will show instead that we obtain the BMS algebra from the invariance of the Hamiltonian under residual gauge transformations in the bulk. 
	
	\vskip 0.2cm
	\ndt
In a previous paper~\cite {Ananth:2020ngt} we have studied the problem by constructing non-linear representations of the Poincar\'e algebra on two field degrees of freedom with helicity $+2$ and $-2$ respectively. We performed the study in the light-cone frame where one of the light-cone coordinates is the evolution parameter ``time'' and the conjugate momentum, the Hamiltonian. The representation is unique up to possible counterterms~\cite{Bengtsson:2012dw}. When we ask if the Hamiltonian is invariant under an extended symmetry we find indeed that it is invariant under the BMS symmetry, which shows that this symmetry is also present in the bulk. It should be mentioned though that our formulation is a perturbative one in $\kappa$, and we only demonstrate this to the lowest order, but experience from earlier studies suggests that the symmetries will hold at higher orders. Furthermore since it is an infinite series with an ever-increasing number of graviton fields we have to restrict our studies to weak fields.

\vskip 0.2cm
	\ndt
The question here is then to ask if the remaining residual gauge invariance  corresponds to just the BMS algebra on the two helicity fields $h$ and $\bar h$ or if it can be further extended. This formulation is often thought to be one where the gauge symmetry is completely fixed apart from some freedom in the definition of the inverse $\partial_-$ (the light-cone momentum that is taken to be a space derivative). But, in this paper, we show that there is still some residual gauge freedom in the  theory, which leads to the light-cone realization of the BMS algebra and only that.
	\vskip 0.2cm
	\ndt The paper is organized as follows. We start with a brief review of the light-cone gauge-fixing of the Einstein-Hilbert action followed by the perturbative expansion in terms of the helicity fields $h$ and $\bar h$. In section 2.2, we discuss the remaining reparameterization freedom in the theory. In section 3, we focus on a special class of reparameterizations, dubbed ``helicity-preserving'', and derive how the fields transform under them by demanding invariance of the light-cone Hamiltonian. We then present the symmetry algebra underlying these reparameterizations, which yields the BMS algebra in light-cone gravity. In section 3.3, we define a canonical generator for the supertranslations, from which we obtain the quadratic form structure for the Hamiltonian previously found in~\cite{Bengtsson:2012dw,Ananth:2017xpj}. We conclude with some remarks on the structure of the BMS algebra in the light-cone gauge and possible extensions to Yang-Mills theory, higher-spin and supersymmetric theories.
	\section{Gravity in the light-cone gauge}
	
	\ndt With the metric $(-,+,+,+)$, the light-cone coordinates are defined as
	\bea
	x^\pm=\fr{\sqrt 2}(x^0\pm x^3)\,,
	\eea
	with the corresponding derivatives $\partial_\pm$. The transverse coordinates and derivatives are
	\bea
	x=\frac{1}{\sqrt 2}\,(\,{x_1}\,+\,i\,{x_2}\,)\ ;&& {\bar\partial} =\frac{1}{\sqrt 2}\,(\,{\partial_1}\,-\,i\,{\partial_2}\,)\, ,  \nn \\
	\bar x=\frac{1}{\sqrt 2}\,(\,{x_1}\,-\,i\,{x_2}\,)\ ;&& {\partial} =\frac{1}{\sqrt 2}\,(\,{\partial_1}\,+\,i\,{\partial_2}\,)\,.
	\eea
	\ndt The Einstein-Hilbert action on a Minkowski background reads 
	\bea
	S_{EH}=\int\,{d^4}x\;\mathcal{L}\,=\,\frac{1}{2\,\kappa^2}\,\int\,{d^4}x\;{\sqrt {-g}}\,\,{\mathcal R}\,,
	\eea
	where $g=det\,(\,{g_{\mu\nu}}\,)$ is the determinant of the metric. $\mathcal R$ is the curvature scalar and $\kappa^2=8\pi G$ is the coupling constant derived from the gravitational constant. The corresponding field equations are
	\bea
	\label{feq}
	\mathcal R_{\mu\nu} \,-\,\half\, g_{\mu\nu}\mathcal R\,=\,0\, .
	\eea
We impose the following {\it {three}} gauge choices~\cite{Scherk:1974zm, BCL} on the dynamical variable $g_{\mu\nu}$
	\bea \label{lcg}
	g_{--}\,=\,g_{-i}\,=\,0\,,\quad i=1,2\, .
	\eea
	\ndt These choices are motivated by the fact that in Minkowski space, we have $\eta_{--}\!=\eta_{-i}\!=0$. We also choose 
	\bea
	\label{gc}
	\begin{split}
		g_{+-}\,&=\,-\,e^\phi\,, \\
		g_{i\,j}\,&=\,e^\psi\,\gamma_{ij}\,.
	\end{split}
	\eea
	where $\phi, \psi$ are real parameters and $\gamma^{ij}$ is a real, symmetric matrix with unit determinant. Field equations that do not involve time derivatives $(\del_+)$ are constraint relations as opposed to {\it true} equations of motion, which have explicit time derivatives. The $\mu\!=\!\nu\!=\!-\;$ constraint from (\ref {feq}) yields
	\bea \label{CE1}
	2\,\del_-\phi\,\del_-\psi\,-\,2\,\del^2_-\psi\,-\,(\del_-\psi)^2\,+\, \half \del_-\gamma^{ij}\,\del_-\gamma_{ij}\,=\,0\, ,
	\eea
	which may be solved by making the {\it {fourth and final}} gauge choice 
	\bea \label{fourth}
	\phi\,=\,\frac{\psi}{2}\,.
	\eea 
	Note that this gauge choice relates $g_{+-}$ and $g_{ij}$. This choice implies, from \eqref{CE1}, that 
	\bea 
	\label{psio}
	\psi\,=\,\frac{1}{4}\,\frac{1}{\del^2_-}\,(\del_-\gamma^{ij}\,\del_-\gamma_{ij})\, .
	\eea 
	Other constraint relations eliminate $g_{++}$ and $g_{+i}$ resulting in the following action 
	\bea
	\label{aaction}
	S\,&=&\frac{1}{2\kappa^2}\int d^{4}x \; e^{\psi}\left(2\,\del_{+}\del_{-}\phi\, +\, \del_+\del_-\psi - \half\,\del_{+}\gamma^{ij}\del_{-}\gamma_{ij}\right) \nonumber \\
	&&-e^{\phi}\gamma^{ij}\left(\del_{i}\del_{j}\phi + \half \del_{i}\phi\del_{j}\phi - \del_{i}\phi\del_{j}\psi - \frac{1}{4}\del_{i}\gamma^{kl}\del_{j}\gamma_{kl} + \half \del_{i}\gamma^{kl}\del_{k}\gamma_{jl}\right) \nn \\
	&&- \half e^{\phi - 2\psi}\gamma^{ij}\frac{1}{\del_{-}}R_{i}\frac{1}{\del_{-}}R_{j}\,,
	\eea
	\ndt where 
	\bea 
	R_{i}\,\equiv\, e^{\psi}\left(\half \del_-\gamma^{jk}\del_{i}\gamma_{jk}-\del_-\del_i\phi - \del_-\del_i\psi + \del_i\phi\del_-\psi\right)+\del_k(e^\psi\,\gamma^{jk}\del_-\gamma_{ij})\,. \nn
	\eea
This is the closed form expression for the light-cone gravity action~\cite{Scherk:1974zm,BCL} - purely in terms of the physical degrees of freedom in the theory.
\vskip 0.3cm
	\subsection{Perturbative expansion}
	\vskip .3cm
	\ndt We now examine the perturbative expansion of the closed form expression obtained above. The order $\kappa^2$ result was first presented in~\cite{BCL,Ananth:2006fh} while the $\kappa^3$ result was derived in~\cite{Ananth:2008ik}. We parameterize the matrix $\gamma_{ij}$ as
	\bea 
	\gamma_{ij}\,=\,(e^{\kappa H})_{ij}\,,
	\eea
	where $H$ is a traceless matrix since $det\,(\,\gamma_{ij})=1$. We choose
	\bea \label{matrixH}
	H\,=\,\begin{pmatrix} h_{11} & h_{12} \\ h_{12} & -h_{11} \end{pmatrix}\,; \quad h\,=\,\frac{(h_{11}+i\,h_{12})}{\sqrt{2}}\,, \,\,\,\,\,\,\,\,\,\,\,\,\, \bar{h}\,=\,\frac{(h_{11}-i\,h_{12})}{\sqrt{2}}\ ,
	\eea
	\ndt The Lagrangian (density) in terms of $h$ and $\bar h$ to order $\kappa$ now reads
	\bea 
	\label{LCLag}
	\mathcal{L}&=& \half \,\bar h\,\Box\, h\,+\, 2\,\kappa\, \bar{h}\, \parm^2\left[-\,h\,\frac{\bar{\del}^2}{\parm^2}h\,+\,\frac{\bar{\del}}{\parm}h\,\frac{\bar{\del}}{\parm}h\right] \,+\, \text{c.c.}\ ,
	\eea
	\ndt \ndt with the d'Alembertian $\Box\,=\,2\,(\,\partial\,{\bar \partial}\,-\,\partial_+\,\parm\,)$. At the next order, time derivatives need to be removed using field redefinitions and the resulting quartic Lagrangian was computed in~\cite{BCL,Ananth:2006fh}. 
	\vskip 0.2cm
	\ndt The corresponding Hamiltonian density for gravity reads 
	\bea \label{LCH}
	\mathcal{H} &=& \del \bar h\, \bar \del h\ +\ 2\, \kappa\, \del_-^2 \bar h\, \left(h\,\frac{{\bar \del}^2}{\del_-^2}h\ -\ \frac{\bar \del}{\del_-}h\, \frac{\bar \del}{\del_-}h\right) + c.c.\ +\ \mathcal O(\kappa^2)\ .
	\eea
	
	\ndt	
	The light-cone action for gravity is invariant under Poincar\'e transformations in four dimensions~\cite{Bengtsson:1983pd}. As $x^+$ is treated as the evolution parameter, the conjugate momentum $P^-$ is the Hamiltonian operator. The light-cone Poincar\'e generators split into two kinds - the kinematical ones ($\mathbb K$) which do not involve time derivatives $\del_{+}$ and the dynamical ones ($\mathbb D$) that do and hence receive non-linear contributions in the interacting theory. The dynamical generators  or ``Hamiltonians'' in Dirac's language~\cite{Dirac} take the field forward in light-cone time.
	\bea \label{K-D split}
	&&\mathbb K:\quad \{P, \bar P, P^+, J, J^+, \bar J^+, J^{+-}\}  \nn \\
	&& \mathbb D:\quad \{P^-\equiv H, J^-, \bar J^-\}
	\eea
	\ndt 	All the relevant commutators of the Poincar\'e algebra in four dimensions are listed in appendix A. The commutators fall broadly into three varieties	
	\bea \label{KDcomm}
	[\, \mathbb K,\, \mathbb K\,]~=~ \mathbb K\, ,& [\,\mathbb K,\, \mathbb D\,]~=~ \mathbb D\,, & [\,\mathbb D,\, \mathbb D\,]~=~ 0\, .
	\eea
	\ndt The fields $h$ and $\bar h$ transform with helicity, $\lambda =2$ and $\lambda =-2$, respectively under the little group in four dimensions and thus, represent the two physical states of graviton. 
\vskip 0.3cm

	\subsection*{The Hamiltonian}
	\ndt 
 The Hamiltonian $H \equiv P^-$ in \eqref{LCH} follows from the action in \eqref{LCLag} -
 which was obtained by gauge-fixing the covariant Einstein-Hilbert
 Lagrangian. There is an alternate approach that also leads to the same
 result - this is to recognize that the Hamiltonian is also an element
 of the Poincaré algebra and hence can be determined entirely simply
 closing all the commutators in the symmetry algebra~\cite{Bengtsson:1983pd}. This is a necessary step because Lorentz invariance is not manifest on the
 light-cone and must be explicitly checked.  It is interesting to note, however, that the Hamiltonian is explicitly helicity-covariant when expressed in terms of $h$ and $\bar h$.
\vskip 0.3cm	
Having reached this point, we note now that there are still some reparameterizations allowed - more specifically, transformations that (i) leave the Hamiltonian invariant and (ii) preserve all gauge choices made so far. These residual reparameterizations will be the focus of the next subsection.

\vskip 0.3cm

	\subsection{Residual gauge transformations}
	
Having obtained a perturbative expansion of the gauge-fixed action, we now turn to examining the issue of residual reparameterization invariace. To this end, we note that under
		\[ x^\mu~ \rightarrow~ x^\mu\ +\ \xi^\mu\,, \]
	\ndt the first gauge choice $g_{--}=0$ in \eqref{lcg} holds as long as 
	\be \label{RRplus}
	\xi^+~=~f(x^+, x^j)\quad \text{so}\quad \del_- \xi^+~=~0\, .
	\ee
	\vskip 0.2cm
	\ndt
	The second gauge condition $g_{-i}=0$ in \eqref{lcg} then requires that 
	\bea \label{RR-i}
	\del_- \xi^j\, g_{ij}\ +\ \del_i \xi^+\, g_{+-} = 0\,.
	\eea
	This relates $\xi^j$ to $\xi^+ = f(x^+, x^j)$
		\be
		\xi^k~=~  - \del_i f\, \frac{1}{\del_-}(g_{+-} g^{ik}) \ + Y^k\,,
		\ee
		\ndt where $Y^k$ does not depend on $x^-$. The fourth gauge condition (\ref{fourth}) further restricts the form of different components of $\xi^\mu$.	
\vskip 0.3cm

\ndt Confining ourselves to residual gauge transformations on the fields $h$, $\bar h$ we look for transformations that leave the light-cone Hamiltonian invariant. The light-cone gauge conditions and the subsequent perturbative expansion of $\gamma_{ij}$ constrain the form of allowed residual reparameterizations, the details of which can be found in appendix B. This choice of residual gauge transformations, when expressed in the $(x, \bar x)$ coordinates, corresponds to the following ``helicity-preserving'' reparameterizations
	\begin{eqnarray} \label{NewRR}
	x & \rightarrow & x\ +\ Y(x)\,,\label{Y} \\
	\bar x &\rightarrow &\bar x\ +\ \overline Y(\bar x)\,, \label{bar Y}\\
	x^+ & \rightarrow& x^+ \, +\ f(x, \bar x, x^+)\,,  \\
	x^- & \rightarrow & x^-\ +\ \xi^-  \,,
	\end{eqnarray}
	such that the parameters satisfy 
	\bea \label{cond}
	\del Y = \del_- Y = 0 \, , \quad \bar  \del \overline Y = \del_- \overline Y = 0\,  , \quad \del_- f = 0\, ,
	\eea
	\ndt and the parameter $\xi^-$ is completely determined in terms of $f, Y, \overline Y$ as in \eqref{xi minus}. 
	
	\section{BMS symmetry in light-cone gravity} 
	
		\subsection{Invariance of the light-cone Hamiltonian}
		Before we compute the symmetry algebra underlying these reparameterizations, we must ensure invariance of the light-cone Hamiltonian expressed in terms of the helicity fields (\ref{LCH}). We must do this since the elimination of the unphysical degrees of freedom might further reduce the residual symmetry. Thus, invariance of the Hamiltonian under the new reparameterizations guarantees that Lorentz covariance is not violated. This requirement will indeed put some additional constraints on the parameters $Y$ and $\overline Y$ as we show below. 
			
			\vskip 0.2cm
			\ndt
			This will, in turn, ensure that we can define a canonical generator, $G_\xi$ for these residual gauge transformations in the reduced phase space of $(h, \bar h)$ such that it commutes with the Hamiltonian
			\be \label{canonical}
			\delta_\xi H ~=~ 0 \quad \Rightarrow \quad  [G_\xi , H]\, ~=~ 0 .
			\ee	
			We can then study the algebra of these residual gauge generators with the canonical Poincar\'e generators defined in the $(h, \bar h)$ phase space, which are presented in Appendix A.
			\vskip 0.2cm
		\ndt 
	One can check the invariance of the light-cone Hamiltonian  under a given transformation order by order in $\kappa$ as follows
		 \bea
		\delta_{\xi} H &=& \delta_{\xi}^{(0)}\, H^{(0)} \,+\,
	 \delta_{\xi}^{(\kappa)}\, H^{(0)}\,+\, \delta_{\xi}^{(0)} \,H^{(\kappa)} \,+\, \mathcal O(\kappa^2)~=~ 0\,.
		\eea
		
		\ndt
		Under the reparameterizations \eqref{Y} and \eqref{bar Y} with the parameters satisfying \eqref{cond}, the fields $h$ and $\bar h$ transform  as
		\bea
		\label{rotation}
		\delta_{Y, \overline Y}\, h&=& Y(x)\, \bar \del h \,+\, \overline Y(\bar x)\, \del h\, +\, (\del \overline Y - \bar \del Y)\,h\,, \ \nn \\
		\delta_{Y, \overline Y}\, \bar h &=& Y(x)\, \bar \del \bar h\, +\, \overline Y(\bar x)\, \del \bar h\,-\, (\del \overline Y - \bar \del Y)\,\bar h \,.
		\eea
		These transformations leave the Hamiltonian invariant if one assumes 
		\be \label{LorentzY}
		{\bar \del}^2 Y = \del^2 \overline Y= 0\, ,
		\ee
		\ndt which restricts the parameters $Y$ and $\overline Y$ to be at most linear in $x$ and $\bar x$ respectively. Thus, the above condition reduces the $Y, \overline Y$ reparameterizations to a part of the Poincar\'e transformations. With these assumptions, it is easy to show that the light-cone Hamiltonian is invariant under (\ref{rotation}) up to the cubic order. More importantly, no new corrections to $\delta_{Y, \overline Y} h$ of order $\kappa$ or higher are required, thus, classifying these under the kinematical part of the light-cone Poincar\'e transformations \eqref{K-D split}.
	
\vskip 0.2cm
\ndt We now examine the time reparameterizations labeled by the parameter $f$, which transforms the fields as
\be
\delta_f h = f\del_{+} h ~=~ f  \frac{\del \bar \del}{\del_-}h\, .
\ee
This transformation leaves the free light-cone Hamiltonian invariant. However, in order to establish the invariance at order $\kappa$
 \bea
\delta_{f}^{(\kappa)}\, H^{(0)}\,+\, \delta_{f}^{(0)} \,H^{(\kappa)} ~=~ 0\,,
\eea
we  must add corrections to $\delta_{f}h$ at $\mathcal{O}(\kappa)$. Thus, the field $h$ transforms non-linearly under $f$ as follows
\bea \label{flocal}
\delta_{f} h&=& f(x, \bar x, x^+)\ \Bigg\{\,\frac{\del \bar \del}{\del_-}h\, +\, 2\,\kappa\,\del_- \left(h\, \frac{{\bar \del}^2}{\del_-^2} h\, -\, \frac{\bar \del}{\del_-}h\, \frac{\bar \del}{\del_-} h\right)\, \nn \\
&&+\,2\,\kappa\, \frac{1}{\del_-^3} 
\left(\frac{\del^2}{\del_-^2}\bar{h}\,\del_-^2 h\, -\, 2\, \frac{\del}{\del_-}\bar{h}\, \del_- \del h\, +\, \bar{h}\,\del_-^2\del^2h\right)+\, \mathcal O(\kappa^2) \Bigg\}\,,
\eea
and $\delta_{f}\bar h$  is simply the complex conjugate of the above expression. Therefore, the above time reparameterizations are a symmetry of the light-cone Hamiltonian, provided one adds corrections to $\delta_{f}h$ at every order in $\kappa$, making these transformations dynamical.	Interestingly, in order to prove the invariance of the Hamiltonian, the explicit form of $f$ is not required.
\vskip 0.2cm
\ndt It is important to note though that {\it the invariance of the Hamiltonian under these transformations is strictly proven only to first order in the coupling constant, $\kappa$}. But, extensions to higher orders should follow without any formal difficulties, although the explicit calculations could prove cumbersome.	
		\vskip 0.2cm
	\ndt A key difference from the previous analysis in~\cite{Ananth:2020ngt} is that the condition $\del_-f =0$ here follows from (\ref{RRplus}), which is a consequence of the light-cone gauge fixing of the Einstein-Hilbert action. In~\cite{Ananth:2020ngt}, the same condition is obtained by demanding the invariance of the light-cone Hamiltonian under local extensions of the Poincar\'e transformations. This reflects how the focus of this paper is primarily on the residual gauge symmetry of the graviton fields $h$ and $\bar h$, in contrary to~\cite{Ananth:2020ngt}, where the goal was to obtain possible extensions of the light-cone Poincar\'e algebra with local parameters.
	
	\subsection{Light-cone realization of the BMS algebra}
	\ndt The fourth gauge condition (\ref{fourth}) precisely fixes the $x^+$ dependence of $f$ as in \eqref{f-appendix}
\bea \label{fBMS}
f(x^+,x,\bar x)&=& T (x, \bar x)\ +\ \frac{1}{2}\, x^+\, (\del \overline Y\ +\ \bar \del Y)\, ,
\eea

\ndt which reproduces the light-cone BMS symmetry in~\cite{Ananth:2020ngt}. Here, the parameters $Y$ and $\overline{Y}$ obey (\ref{cond}) and (\ref{LorentzY}), indicating that the only independent parameter in $f$ is $T(x, \bar x)$. Besides, this choice of $f$ also coincides with the conformal Carroll case as we have discussed in appendix C. As opposed to the geometric aspects of the conformal Carroll group in~\cite{Duval:2014uva}, our focus lies on the field-theoretic aspects of gravity and the Lie algebra representation associated with the symmetries of the theory. 

\vskip 0.2cm
\ndt On the initial surface $x^+=0$ such that $f=T$, the BMS transformations in light-cone gravity to order $\kappa$ read
	\bea 
	\delta_{Y, \overline Y, T}\, h&=& Y(x)\, \bar \del h + \overline Y(\bar x)\, \del h\ + (\del \overline Y - \bar \del Y)\,h + T\,\frac{\del \bar \del}{\del_-}\,h  \nn \\
	&&-\, 2\, \kappa\, T\, \del_-\, \left(h\, \frac{{\bar \del}^2}{\del_-^2} h\, -\, \frac{\bar \del}{\del_-}h\, \frac{\bar \del}{\del_-} h\right)  \ -\ 2\, \kappa\, T\, \frac{1}{\del_-}\left(\frac{{\del}^2}{\del_-^2}\bar h\, \del_-^2 h\right)\, \nn \\
	&& -\, 2\, \kappa\, T\, \frac{{\del}^2}{\del_-^3}\, (\bar h\, \del_-^2  h)\, +\ 4\, \kappa\, T\, \frac{\del}{\del_-^2} \left(\frac{\del}{\del_-}\bar h\, \del_-^2 h \right)\, ,
	\eea 
	where the parameters $Y$, $\overline Y$ and $T$ satisfy
	\be \label{cond1}
	\del Y = \del_- Y = 0 \, , \quad \bar  \del \overline Y = \del_- \overline Y = 0\,, \quad \del_- T = 0\,,
	\ee
	and 
	\be \label{cond2}
	\bar{\del}^2Y~=~\del^2 \overline Y~=~0\, .
	\ee
	Two such transformations close on another reparameterization	
	\bea \label{alg}
	\left[\, \delta(Y_1, \overline Y_1, T_1)\, ,\ \delta (Y_2, \overline Y_2, T_2) \, \right]\, h &=& \ \delta (Y_{12}, \overline Y_{12}, T_{12})\, h\,,
	\eea
	\ndt with the new parameters defined as
	\bea
	Y_{12}& \equiv& Y_2\, \bar \del\, Y_1\ -\ Y_1\, \bar \del\, Y_2\,, \\
	\overline{Y}_{12} & \equiv & \overline Y_2\, \del\, \overline Y_1\ - \ \overline Y_1\, \del\, \overline Y_2\,, \\
	T_{12} & \equiv & [Y_2\, \bar \del \, T_1\ +\ \overline{Y_2}\, \del\, T_1\ +  \frac{1}{2}\, T_2 (\bar \del Y_1\ +\, \del \overline Y_1) ]\ - \ (1 \leftrightarrow 2)\, .
	\eea
	\ndt This is the light-cone realization of the BMS algebra in four dimensions~\cite{Ananth:2020ngt}. The ``supertranslations'' labeled by the function $T(x, \bar x)$ enhance the dynamical part of the Poincar\'e algebra into an infinite-dimensional set, while the kinematical part of the algebra is the same as in (\ref{K-D split})
	\bea \label{k-T split}
	&&\mathbb K~\rightarrow~\mathbb K\,, \nn \\
	&& \mathbb D~\rightarrow~ \mathbb D (T)\,,
	\eea
	\ndt
	such that the light-cone  BMS algebra in (\ref{alg}) takes the form
		\bea \label{LC-BMS}
	[\, \mathbb K,\, \mathbb K\,]~=~ \mathbb K\, ,& [\,\mathbb K,\, \mathbb D (T)\,]~=~ \mathbb D (T)\,, & [\,\mathbb D (T),\, \mathbb D (T)\,]~=~ 0\,. 	\eea
	
\ndt Thus, the key feature of the light-cone BMS algebra is that the dynamical part of the Poincar\'e algebra is enlarged to accommodate the supertranslations. This is different from the BMS algebra in covariant formulations, where the four spacetime translations of the Poincar\'e algebra are enhanced by the supertranslations. However, Lorentz invariance of the theory dictates that the enhancement involves \textit{only one single local parameter} $T(x,\bar x)$, which is in keeping with the original BMS group found in~\cite{Bondi:1962px, Sachs:1962zza}.
\vskip 0.2cm
\ndt The Poincar\'e subgroup of the BMS algebra can be obtained by imposing the condition $\del^2T= \bar{\del}^2 T=0$. This restricts $T$ to be at most linear in $x$ or $\bar x$, thereby, reducing the dynamical part to $\mathbb D$ given in (\ref{K-D split}). In appendix B, we discuss in more details the Poincar\'e subgroup of the light-cone BMS algebra \eqref{Poincare in BMS}.

	\subsection{Supertranslations and the quadratic form Hamiltonian}
	
	As we have established that $\delta_T h$ are canonical transformations in the $(h, \bar h)$ phase space, we can, now, define a canonical generator $G_T$ for supertranslations through \eqref{canonical}
	\bea
	G_T&=& \int d^3x\, \del_- \bar h \, (\delta_T h)~=~\int\, d^3x\, \del_- \bar{h}\, T\, \frac{\del \bar \del }{\del_-}h\, +\, \mathcal O(\kappa)\,,
	\eea
	\ndt where we have used the equation of motion for $h$ obtained from (\ref{LCLag}). One can derive the transformation law for the fields $h$ and $\bar{h}$ order by order in $\kappa$ from the above generator using the brackets{\footnote{The fields $h$ and $\bar{h}$ satisfy \be[h(x), \bar{h}(y)]\, =\, \frac{1}{\del_-} \delta^{(3)} (x-y)\, ,\quad [\, h(x), h(y)\,]\, =\, [\,\bar h(x), \bar{h}(y)\,]\, =\, 0\, .\ee}}
		\bea
	\delta_T h~=~ [G_T, h]\,, && \delta_T\bar{h}~=~ [G_T, \bar{h}]\, .
	\eea
	\ndt At the lowest order, the generator can be brought to the form
	\bea \label{STgen0}
	G_T^{(0)}&=& \int\, d^3x\, \left(T\, \del \bar{h}\, \bar{\del}h\, +\, \del T\, \bar{h}\, \bar{\del}h \right)\ \nn \\
	&=&\int d^3x\, \left(T \, \del \bar{h}\, \bar{\del}h+ \frac{1}{2} \del T\, \bar h\, \bar{\del}h + \frac{1}{2}\, \bar{\del}T\, h\, \del \bar{h} \right)\,.
	\eea
	\ndt At the cubic order, $G_T$ reads~\cite{Bengtsson:1983pd}
	\bea
	G_T^{(\kappa)}&=&\kappa \int d^3x\, T\,\del_- \bar{h}\Bigg\{2\,\del_- \left(h\, \frac{{\bar \del}^2}{\del_-^2} h\, -\, \frac{\bar \del}{\del_-}h\, \frac{\bar \del}{\del_-} h\right)\, \nn \\
	&&+\, \frac{1}{\del_-^3} 
	\left(\frac{\del^2}{\del_-^2}\bar{h}\,\del_-^2 h\, -\, 2\, \frac{\del}{\del_-}\bar{h}\, \del_- \del h\, +\, \bar{h}\,\del_-^2\del^2h  \right) \Bigg\}\,.  
	\eea
	After some partial integrations, the details of which can be found in Appendix \ref{sec:B}, the supertranslation generator up to order $\kappa$ reads
	\bea \label{STgen}
	G_T&=& \int d^3x\, T\, \mathcal D\bar{h}\, \overline{\mathcal D}h\, +\, \,\int d^3x\,\Bigg\{\frac{1}{2}\,\del T\, \bar h\, \bar{\del}h\,-\,\kappa\,\del T\,\bar h\frac{1}{\del_-^2}\left(\frac{\del}{\del_-}\bar{h}\, \del_-^3 h- \bar h\del_-^2 \del h\right) \nn \\
	&-&\,2\, \kappa\,\del T\,\frac{1}{\del_-^2}(\bar h\del_-^2 h)\, {\del} \bar h \,+\,\frac{1}{2}\, \kappa\,\del^2 T\, \bar{h}\, \frac{1}{\del_-^2}(\bar{h}\del_-^2 h)\,\Bigg\}+\ \text{c.c.}\, ,
	\eea 
	where the ``covariant'' derivative $\overline{\mathcal D} h$ reads~\cite{Bengtsson:2012dw,Ananth:2017xpj}
	\bea
	\overline{\mathcal D} h&=& \bar{\del}h\, +\, 2\, \kappa\, \frac{1}{\del_-^2}\left(\frac{\del}{\del_-}\bar{h}\,\del_-^3h\,-\, \bar{h}\del_-^2\del h\right)\, .
	\eea

	\ndt For constant time translations, $T=a$, the corresponding generator becomes the Hamiltonian of the theory given by
	\bea \label{QFH}
	G_{T=a}~=~ H~=~\int d^3x\, a\, \mathcal D\bar{h}\, \overline{\mathcal D}h\, .
	\eea
	Thus, the light-cone Hamiltonian for gravity in four dimensions can be recast into a \textit{positive semi-definite structure}, which we call \textit{quadratic form}, indicating that the energy of the system is manifestly positive. In an earlier work~\cite{Ananth:2017xpj}, we had derived the order $\kappa^2$ terms in $\bar{\mathcal D} h$ proving that this feature extends to the second order in coupling constant as well. However, in this analysis, we learn that the quadratic form structure of the Hamiltonian is actually fixed by the BMS symmetry. The terms involving derivatives of $T$ in (\ref{STgen}) are not arbitrary as these are related to the \textit{spin corrections} to the boost generators $J^-$ and $\bar J^-$ required for the closure of the light-cone Poincar\'e algebra. Any further partial integrations in $\del$ and $\bar{\del} $ will spoil this structure and hence, lead to ill-defined boost generators\footnote{This can be compared to the asymptotic symmetry analysis at spatial infinity, where the boost generators require special treatment in order to have well-defined Poincar\'e generators in the Hamiltonian formulation~\cite{Henneaux:2018cst, Henneaux:2019yax}.}. In fact, we can read off corrections to the boost generators order by order in $\kappa$ from the general expression for supertranslation generator to higher orders. 
	\vskip 0.2cm
	\ndt
	The quadratic form expression (\ref{QFH}) resembles the generator of supertranslations at null infinity  proposed by Bondi, van der Burg, Metzner, and Sachs~\cite{Bondi:1962px,Sachs:1962zza},
	which shows that the flux of the gravitational energy at null infinity is positive. Proof of the positive energy theorem in gravity~\cite{Deser:1977hu,Witten:1981mf,Ashtekar:1986yd} typically involves some spinor-like variables. In the light-cone formulation, however, one can derive the quadratic form Hamiltonian without resorting to any spinor variables. It is important to note though that we are concerned with the energy density in the bulk, not the flux of gravitational energy at null infinity.
	\vskip 0.2cm
	
	\ndt
	Interestingly, the covariant derivative $\overline{\mathcal D} h$ in the quadratic form Hamiltonian plays a role similar to the ``News tensor'' in the Bondi frame, which is related to vacuum configurations. Therefore, we can define the vacuum configurations of gravity in the light-cone formulation as the states which satisfy the condition
		\bea
		\overline{\mathcal D} h&=& \bar{\del}h\, +\, 2\, \kappa\, \frac{1}{\del_-^2}\left(\frac{\del}{\del_-}\bar{h}\,\del_-^3h\,-\, \bar{h}\del_-^2\del h\right)\, +\mathcal O(\kappa^2)~=~0\, .
		\eea
	\ndt 
	It would be instructive to use this equation to classify different vacuum configurations and explore its possible implications for the quantum theory, as the fields $h$ and $\bar h$ indeed correspond to the two physical states of the graviton.

	\section{Concluding remarks}
	\ndt In this paper, we discussed the BMS symmetry in the light-cone gauge from the perspective of residual gauge freedom in the theory. The invariance of the Hamiltonian under these reparameterizations leads to the correct transformation laws for the fields, which then realize the BMS algebra in four dimensions. The most notable characteristics of the light-cone BMS algebra is the enhancement of the dynamical part of the Poincar\'e algebra. In the light-cone formalism, the dynamical transformations, which are non-linearly realized on the fields, provide us with a powerful framework to construct interacting actions from the closure of the symmetry algebra~\cite{Bengtsson:1983pd}.  This is particularly interesting for higher-spin theories as constructing interacting actions with higher-spin fields is a very complex issue. It would, therefore, be worthwhile to explore the implications of these non-linear transformations for the interacting theory. 

\vskip 0.2cm
\ndt Although ours is a perturbative analysis, this light-cone approach offers a unique insight into the structure of the BMS symmetry, that also shares some interesting features with the BMS literature for the full Einstein theory. One important similarity with the spatial infinity analysis is that the invariance of the Hamiltonian reduces the parameters, $Y$ and $\overline Y$, to Lorentz rotations, which eliminates superrotations from the theory.	In light-cone gravity, one can study the BMS symmetry in a physical gauge, which might help us better understand its connection with on-shell amplitudes~\cite{SAMHV1,SAMHV2} and soft theorems~\cite{Strominger:2014pwa, Donnay:2018neh}.

	\vskip 0.2cm
	\ndt The fact that the light-cone Hamiltonian for gravity in four dimensions can be expressed as a positive semi-definite quadratic form puts this theory in a special class of theories that admit such Hamiltonians. The occurrence of such simple structures in field theories is often indicative of hidden symmetries~\cite{ABM1,ABM2}. In this paper, we have attributed the quadratic form structure in light-cone gravity to the invariance of the Hamiltonian under BMS supertranslations. Similar analyses for residual gauge symmetry in Yang-Mills and higher-spin theories, which also admit such quadratic form Hamiltonians~\cite{Ananth:2015tsa,Ananth:2020mws}, could bring forth some deeper links between these theories and gravity. Similarly, extensions to supersymmetric theories~\cite{Awada:1985by} could be instrumental in explaining the quadratic form Hamiltonians found in maximally supersymmetric Yang-Mills and supergravity in four dimensions.

	\section*{Acknowledgments}
	We thank Glenn Barnich and Marc Henneaux for many discussions. The work of SM is partially supported by the ERC Advanced Grant ``High-Spin-Grav'',  by FNRS-Belgium (conventions FRFC PDRT.1025.14 and  IISN 4.4503.15), as well as by funds from the Solvay Family.
	\appendix

	\section{Light-cone Poincar\'e algebra in $d=4$}
	In this section, we present the light-cone Poincar\'e algebra in four dimensions. 
	A general Poincar\'e trasnformation in four dimensions
		\be
		\delta x^\mu ~ = ~ \omega^\mu_\nu x^\nu + a^\mu \,,
		\ee
		in light-cone coordinates, takes the unusual form
		\be \label{LC poincare}
		\delta \begin{pmatrix}
			x^+ \\ x^- \\ x \\  \bar{x}
		\end{pmatrix} ~ = ~  \begin{pmatrix}
			\omega_{+-} & 0 & - \bar{\omega}_- & -\omega_- \\
			0& -\omega_{+-} & \bar{\omega}_+ & \omega_+ \\
			\omega_+ & \bar{\omega_-} & -i \omega_{12} &0 \\
			\bar{\omega_+} & \omega_- & 0 & i \omega_{12}
		\end{pmatrix} \begin{pmatrix}
			x^+ \\ x^- \\ x \\ \bar{x}
		\end{pmatrix} \ +\ \begin{pmatrix}
			a^+ \\ a^- \\ a \\  \bar{a}
		\end{pmatrix}
		\ee	
		where the two real $(\omega_{+-}, \omega_{12})$ and two complex $(\omega_+, \omega_-)$ parameters label the Lorentz group.
	We define 	
	\bea
	J^+\ =\ \frac{J^{+1}+iJ^{+2}}{\sqrt{2}}\ ,\quad \bar J^{+}\ =\ \frac{J^{+1}-iJ^{+2}}{\sqrt 2}\ ,\quad J\ =\ J^{12}\ ,\quad H\ =\ P^-\ .
	\eea
	\vskip 0.2cm
	\ndt
	All the {\it {non-vanishing}} commutators of the Poincar\'e algebra are listed below
	\bea
	&[H, J^{+-}]\ =\ -i H \ , \quad  &[H, J^+] = -i P\ , \qquad \qquad \quad [H, \bar J^+]\ =\ -i \bar P \nn \\ \nn\\
	&[P^+, J^{+-}]\ = \ iP^+\ ,\quad  &[P^+, J^-]\ =\ -iP\ , \qquad \qquad [P^+, \bar J^-]\ =\ -i \bar P \nn \\ \nn \\
	&[P, \bar J^-]\ =\ -i H\ ,  \quad &[P, \bar J^+] \ =\ -iP^+\ , \qquad \qquad [P, J]\ =\ P \nn \\ \nn \\
	&[\bar P, J^-]\ = \ -i H\ , \quad &[\bar P, J^+]\ =\ -iP^+\ , \qquad \qquad [\bar P, J]\ =\ -\bar P \nn \\ \nn \\
	&[J^-, J^{+-}]\ =\ -i J^- \ , \quad &[J^-, \bar J^+]\ =\ iJ^{+-} +  J \ , \qquad [J^-, J]\ =\ J^- \nn \\ \nn \\
	&[\bar J^-, J^{+-}]\ = \ -i \bar J^- \ , \quad &[\bar J^-, J^+]\ =\ iJ^{+-} - J \ , \qquad [\bar J^-, J]\ =\ -\bar J^- \nn \\ \nn \\
	&[J^{+-}, J^+]\ =\ -i J^+ \ ,  \quad &[J^{+-}, \bar J^+]\ = \ -i \bar J^+ \ , \nn \\ \nn \\
	&[J^+, J]\ =\ J^+\ , \quad &[\bar J^+, J]\ =\ -\bar J^+\ .
	\eea
	
	\vskip 0.3cm
	\ndt
	In case of light-cone gravity, the Poincar\'e generators are canonically realised on the phase space of $h$ and $\bar h$ as follows
		
	\bea
	&&P^-\, =\, \int d^3x\, \mathcal H\ , \quad P \,=\, \int d^3x \del_-\bar h\, \del h\,, \quad \overline P \,=\, \int d^3x \del_- \bar h\, \bar \del h\,, \quad P^+ \,=\,\int d^3x \del_{-} \bar h\, \del_- h\,, \nn \\ 
	&&J\,=\, i\,\int d^3x\, \del_- \bar h\, (x \bar \del - \bar x \del - \lambda)\, h\ , \quad J^{+-}\,=\, \int d^3x\, (\del_- \bar h\, x^- \del_- h -x^+ \mathcal{H})\,, \nn \\
	 &&J^+ \,=\, \int d^3x\, \del_{-} \bar h\,(x^+ \del + x \del_{-})\,h\ ,\quad J^- \,=\,\int d^3x\, \left\{x\mathcal{H} + \del_{-} \bar h\left(x^- \del - \lambda \frac{\del}{\del_{-}}\right)h + \mathcal{S}^-\right\}\,, \nn \\
	 &&\bar{J}^+ \,=\, \int d^3x\, \del_{-} \bar h\, (x^+ \bar \del + \bar x\del_{-})\, h\ ,\quad \bar{J}^- \,=\, \int d^3x\, \left\{ \bar x \mathcal{H}+ \del_{-} \bar h\, \left(x^- \bar \del - \lambda \frac{\bar \del}{\del_-}\right)\,h + \bar{\mathcal{S}}^-\right\}\,, \nn \\
	\eea
	where the spin corrections read
	\bea
	\delta_{s^-} h &=&-2\, \frac{\del}{\del_{-}} h\,  - 4\,\kappa\, \del_{-}\left(h\frac{\bar{\del}}{\del_-^2}h - \frac{1}{\del_{-}}h\, \frac{\bar{\del}}{\del_{-}}h\right)\, +\, \mathcal{O}(\kappa^2) \,,\nn \\
	\delta_{\bar{s}^-}h &=& 2\, \frac{\bar{\del}}{\del_{-}} h\, - 4\, \kappa \, \frac{1}{\del_{-}^3} \left(\frac{\del}{\del_{-}^2}\bar h \del_{-}^4 h - \frac{1}{\del_{-}} \bar{h} \del_{-}^3 \del h + 3 \frac{\del }{\del_{-}}\bar h \del_{-}^3 h - 3 \bar h \del_{-}^2 \del h \right) \, +\, \mathcal{O}(\kappa^2) \,.\nn \\
	\eea

\section{Allowed residual gauge transformations}

\ndt Any residual gauge transformation must leave the light-cone gauge choices invariant. These put some constraints on the form of the allowed gauge parameters $\xi^\mu$. 
	
	\vskip 0.2cm
	\ndt The first gauge condition $g_{--} = 0$ leads to the constraint
		\bea\label{cond1}
		\delta g_{--} = 0  &\Rightarrow& \del_- \xi^+\, g_{+-}~=~0  \eea
		\ndt This condition is easily satisfied if we chose the parameter $\xi^+$ as
		
		\[\xi^+~=~f(x^+, x^j)\quad \text{such that}\quad \del_- f~=~0\]
		\vskip 0.2cm
		\ndt
		The condition $g_{-i} = 0$ leads to
		\bea \label{cond2}
		\delta g_{-i} = 0 &\Rightarrow& \del_- \xi^j\, g_{ij}\ +\ \del_i \xi^+\, g_{+-} = 0
		\eea
		
		\ndt For the second condition to hold, we can solve for $\xi^j$ in terms of $\xi^+ = f$
		\be
		\xi^k~=~  - \del_i f\, \frac{1}{\del_-}(g_{+-} g^{ik}) \ + Y^k\ \ ,
		\ee

\ndt where the integration constant $Y^k$ does not depend on $x^-$. 
\vskip 0.2cm
\ndt The gauge condition (\ref{fourth}) relates the $g_{-+}$ component to the determinant of the metric $g_{ij}$. Thus, we first consider $\delta g_{-+}$ to obtain
\be \label{delta-phi}
\delta \phi ~=~ f\, \del_{+}\phi + \xi^- \del_{-} \phi + \xi^k\, \del_k \phi + \del_{+} f + \del_{-} \xi^- - \del_k f\, g_{+i}\,g^{ik}
\ee
As evident from the expression of $\phi$ in (\ref{psio}), the above equation is complicated involving the non-local $\frac{1}{\del_-}$ operators. But in the perturbative expansion, one can simplify the equation in orders of $\kappa$ and obtain non-trivial relations between the gauge parameters. At the lowest order, one finds 
\be \label{delta phi lowest}
\del_{+} f + \del_- \xi^- ~=~ 0
\ee
\ndt 
We can, alternatively, obtain $\delta \phi$ from the variation of the determinant of $g_{ij}$
\be
\delta g = g  \, g^{ij} \delta g_{ij} \,,
\ee
with $g$ given by
\be
g = det (g_{ij}) = 2 \psi = \phi
\ee
\ndt 
where the last equality follows from the fourth gauge choice (\ref{fourth}). When compared with $\delta \phi$ in (\ref{delta-phi}) to the lowest order, we get
\be
\del_{+}f = \frac{1}{2} \del_i Y^i \, .
\ee 
This constraint fixes the time dependence of the parameter $f$ 
\be
f(x^+, x^i) = \frac{1}{2}\del_{i} Y^i\, x^+ + T(x^i) \, ,
\ee
which in the $(x, \bar x)$ coordinate reads
\be \label{f-appendix}
f(x^+, x, \bar x) = \frac{1}{2} x^+ (\del \overline{Y} + \bar{\del} Y)  + T(x, \bar x)\, .
\ee
Further consistency checks may be performed on the remaining components of the metric. Since both $g_{++}$ and $g_{+i}$ are at least of order $\kappa$, we have the following conditions at the zeroth order
\bea
\delta g_{++} = 0 &\Rightarrow& \del_{+} \xi^- = 0\,, \\
\delta g_{+i} = 0 &\Rightarrow& \del_{i} \xi^- = \del_{+} \xi_i \, ,
\eea
which along with \eqref{delta phi lowest} completely determines the form of $\xi^-$ in terms of $f$ and $\xi^i$
\be \label{xi minus}
\xi^- = - (\del_{+} f)\, x^- + (\del_{+}\xi_i)\, x^i\, .
\ee
\subsection*{Poincar\'e subgroup within the BMS}
From the above conditions, one can write a general BMS transformation on the light-cone coordinates as
\bea
\delta x &=& \alpha\, +\, \beta x\, +\, \sigma x^+ \,,\\
\delta \bar{x} &=& \bar{\alpha}\, +\, \bar{\beta} \bar x\, + \, \bar{\sigma} x^+ \,, \\
\delta x^+ &=& Re(\beta) x^+ \, + \, T(x,\bar{x}) \nn \\
&=& Re(\beta)x^+ \, +\, [\,t_0 + t_1 x + \bar{t_1} \bar{x} + \ldots ] \label{T exp}\,,\\
\delta x^- &=& \gamma \, -\, Re(\beta) x^-\,+\, \sigma x\, +\, \bar{\sigma} \bar{x}\,,
\eea
where the parameters $t_0, \gamma$ are real and $\alpha, \beta, \sigma, t_1$ are all complex.
We can then read off the Poincar\'e subgroup inside the BMS group by identifying the parameters in the above set of equations to those in \eqref{LC poincare} as follows
\bea \label{Poincare in BMS}
&&(a^+, a^-, a, \bar{a})~ =~ (t_0, \gamma, \alpha, \bar{\alpha}) \nn \\
&&\omega_{+-} = Re(\beta)\,,\quad \omega_{12} = Im(\beta)\,, \quad (\omega_+, \omega_-)~=~ (\sigma, t_1)\,.
\eea
The extension of the Poincar\'e group to the BMS group is, thus, parameterized \textit{only} by the higher order expansions of $T(x, \bar x)$ in \eqref{T exp}.

	\section{Connections to the Carroll group}
	
	\ndt For null hypersurfaces like the light-cone surfaces $x^{\pm}= constant$, the BMS algebra in four dimensions is isomorphic to the conformal Carroll group~\cite{Duval:2014uva}. The conformal Carroll group consists of two-dimensional conformal algebra on the spatial coordinates augmented by supertranslations in the time direction, $u$.
	\bea
	x\ \rightarrow \phi (x)\ ,\ \quad u\ \rightarrow \Omega(x)^{\frac{1}{2}}\, [u\ +\ \alpha (x)]\,,
	\eea
	where $\Omega(x)$ is the scaling factor associated with the conformal transformations on $x$.
	
	\vskip 0.2cm
	\ndt We, therefore, consider conformal Carroll transformations in the light-cone coordinates on a constant $x^-$ surface
	\bea 
	&&x ~\rightarrow~ \omega(x)\ , \quad \bar x ~\rightarrow~ \overline \omega(\bar x)\,, \\[0.2cm]
	&&x^+~ \rightarrow~ \Omega (x, \bar x)^{\frac{1}{2}}\, [\,x^+\ +\ \alpha ( x, \bar x)\, ]\, \label{LC-CC}.
	\eea
	\vskip 0.2cm
	\ndt
	An infinitesimal conformal transformation on $x, \bar x$ is given by
	\bea \label{GCT}
	x & \rightarrow & \omega(x)\ =\ x\ +\ Y(x)\ , \\ [0.2cm]
	\bar x &\rightarrow &\overline \omega(\bar x)\ =\ \bar x\ +\ \overline Y(\bar x)\ .
	\eea
	\ndt 
	Thus, from (\ref{LC-CC}), we find that the time-coordinate $x^+$ transforms infinitesimally as
	\bea
	x^+ & \rightarrow&  \Omega(x, \bar x)^{\frac{1}{2}}\, \left[ x^+ \, +\  \alpha ( x, \bar x)\right]~=~ [1+ (\bar \del Y +\del \overline{Y})]^{\frac{1}{2}}\, \left(x^+ \, +\  \alpha ( x, \bar x)\right) \nn \\ [0.3cm]
	& \sim& \left\{1+ \frac{1}{2}(\bar \del Y +\del \overline{Y})\right\} \left(x^+ \, +\  \alpha ( x, \bar x)\right) ~\sim~ x^+ \ +\ f(x^+, x, \bar x)\,,
\end{eqnarray}
where
\bea \label{f cond}
f (x^+, x, \bar x)&=&T(x, \bar x)\ +\ \frac{1}{2}\, x^+\, (\bar \del Y\ +\, \del \overline Y)\, .
\eea
\ndt
Here we have ignored the terms of higher orders in $Y$ and $\overline Y$. Thus, the residual reparameterizations considered in (\ref{fBMS}) can indeed be interpreted as infinitesimal conformal Carroll transformation on the constant $x^-$ null hypersurface.

	 \section{Supertranslation generator at order $\kappa$} \label{sec:B}
	 
	 We consider the supertranslation generator at order $\kappa$
	 \bea
	 G_T^{(\kappa)}&=&\kappa \int d^3x\, T\,\del_- \bar{h}\Bigg\{2\,\del_- \left(h\, \frac{{\bar \del}^2}{\del_-^2} h\, -\, \frac{\bar \del}{\del_-}h\, \frac{\bar \del}{\del_-} h\right)\,\nn \\
	 && \quad +\, \frac{1}{\del_-^3} 
	 \left(\frac{\del^2}{\del_-^2}\bar{h}\,\del_-^2 h\, -\, 2\, \frac{\del}{\del_-}\bar{h}\, \del_- \del h\, +\, \bar{h}\,\del_-^2\del^2h  \right) \Bigg\} \nn \\ [0.2cm]
	 &=& \mathcal X + \mathcal Y\, ,
	 \eea
	 \ndt
	 where $\mathcal X$ contains the terms involving $\bar{h}hh$
	 \bea \label{X}
	 \mathcal X&=&2\,\kappa \int d^3x\, T\,\del_- \bar{h}\, \del_- \left(h\, \frac{{\bar \del}^2}{\del_-^2} h\, -\, \frac{\bar \del}{\del_-}h\, \frac{\bar \del}{\del_-} h\right)\, ,
	 \eea
	 \ndt and $\mathcal Y$ contains terms involving $\bar h \bar h h$
	 \bea
	 \mathcal Y&=&\kappa \int d^3x\, T\,\del_- \bar{h}\frac{1}{\del_-^3} 
	 \left(\frac{\del^2}{\del_-^2}\bar{h}\,\del_-^2 h\, -\, 2\, \frac{\del}{\del_-}\bar{h}\, \del_- \del h\, +\, \bar{h}\,\del_-^2\del^2h  \right)\, .
	 \eea
	 \ndt Let us focus on the $\mathcal X$ terms, which upon partial integrations, can be expressed as
	 \bea
	 \mathcal X&=& 2\kappa\int\, d^3x\, T \del_-\bar{h}\, \del_-\left(h\, \frac{{\bar \del}^2}{\del_-^2} h\right)\, -\,2\kappa\int\, d^3x\, T \del_-\bar{h}\, \del_-\left(\, \frac{\bar \del}{\del_-}h\, \frac{\bar \del}{\del_-} h\right) \\
	 &=& -\, 2\kappa\int\, d^3x\,\frac{\bar{\del}}{\del_-^2}(T\, \del_-^2 \bar{h}h)\, \bar{\del}h + 2\kappa\int\, d^3x\,T\, \frac{1}{\del_-}\left(\frac{\bar{\del}}{\del_-}h\,\del_-^2 \bar{h}\right) \\
	 &=& -\, 2\kappa\int\, d^3x\, \bar{\del}T\, \frac{1}{\del_-^2}(\del_-^2 \bar{h}h)\, \bar{\del}h -\, 2\kappa\int\, d^3x\,T\,\frac{\bar{\del}}{\del_-^2}(\del_-^2 \bar{h}h)\, \bar{\del}h \nn \\
	 && + 2\kappa\int\, d^3x\,T\, \frac{1}{\del_-}\left(\bar{\del}h\,\del_-^2 \bar{h}\,+\, \frac{\bar{\del}}{\del_-}h\,\del_-^3 \bar{h}\right) \\
	 &=&  - 2\,\kappa\int d^3x\, \bar{\del}T\, \frac{1}{\del_-^2}(h\del_-^2 \bar h)\, \bar{\del} h \,+\,2\,\kappa \int d^3x\, T\, \frac{1}{\del_-^2}\left(\frac{\bar\del}{\del_-}h\,\del_-^3\bar{h}\,-\, h\,\del_-^2 \bar \del \bar h\right)\bar{\del}h\,.\qquad \quad 
	 \eea
	 \ndt Similarly, the $\mathcal Y$ terms can be simplified as follows
	 \bea
	 \mathcal Y&=&\kappa \int d^3x\,T\,\del_- \bar{h}\frac{1}{\del_-^3} 
	 \left(\frac{\del^2}{\del_-^2}\bar{h} \del_-^4 h  -  2  \frac{\del}{\del_-}\bar{h}  \del_-^3 \del h  +  \bar{h} \del_-^2\del^2h  \right)\\
	 &=& -\kappa \int d^3x\,T\,\del_-^2 \left(\frac{1}{\del_-^2} \bar h \frac{\del^2}{\del_-^2}\bar{h}\right)\del_-^2 h  +  2 \kappa \int d^3x\,\del_- \del \left(T\frac{1}{\del_-^2} \bar h  \frac{\del}{\del_-}\bar{h}\right)  \del_-^2h \qquad \nn \\
	 &&  -  \kappa \int d^3x\,\del^2\left(T\,  \bar{h}\frac{1}{\del_-^2}\bar{h}\right) \del_-^2h  \\
	 &=& -2\kappa\int d^3x \, \del T\, \bar h\frac{1}{\del_-^2}\left(\frac{\del}{\del_-}\bar{h}\, \del_-^3 h- \bar h\del_-^2 \del h\right) + \kappa\int d^3x\, \del^2 T\, \bar{h}\, \frac{1}{\del_-^2}(\bar{h}\del_-^2 h)\qquad \nn \\
	 && -2\kappa\int d^3x\, T \,\del_-^2h\left(\bar h\frac{\del^2}{\del_-^2}\bar h- \frac{\del}{\del_-}\bar h\, \frac{\del}{\del_-}\bar h\right) .
	 \eea
	 \ndt Note that the second line in the above equation is the complex conjugate of $\mathcal X$ in (\ref{X}). Thus, we obtain
	 \bea
	 \mathcal Y&=&-2\kappa\int d^3x \, \del T\, \bar h\frac{1}{\del_-^2}\left(\frac{\del}{\del_-}\bar{h}\, \del_-^3 h- \bar h\del_-^2 \del h\right) + \kappa\int d^3x\, \del^2 T\, \bar{h}\, \frac{1}{\del_-^2}(\bar{h}\del_-^2 h)  \nn \\
	 && - 2\,\kappa\int d^3x\, {\del}T\, \frac{1}{\del_-^2}(\bar h\del_-^2 h)\, {\del} \bar h \,+\,2\,\kappa \int d^3x\, T\, \frac{1}{\del_-^2}\left(\frac{\del}{\del_-}\bar h\,\del_-^3{h}\,-\, \bar h\,\del_-^2 \del h\right){\del}\bar h\, . \qquad \quad 
	 \eea


\begin{thebibliography}{Ref}
		
		\bibitem{Bondi:1962px}
		H.~Bondi, M.~G.~J.~van der Burg and A.~W.~K.~Metzner,
		``Gravitational waves in general relativity. 7. Waves from axisymmetric isolated systems,''
		Proc.\ Roy.\ Soc.\ Lond.\ A {\bf 269} (1962) 21.
			\vskip 0.1cm
		R.~K.~Sachs,
		``Gravitational waves in general relativity. 8. Waves in asymptotically flat space-times,''
		Proc.\ Roy.\ Soc.\ Lond.\ A {\bf 270} (1962) 103.
			
		\bibitem{Sachs:1962zza}
		R.~Sachs,
		``Asymptotic symmetries in gravitational theory,''
		Phys.\ Rev.\  {\bf 128} (1962) 2851.
		\bibitem{Barnich:2009se} 
		G.~Barnich and C.~Troessaert,
		``Symmetries of asymptotically flat 4 dimensional spacetimes at null infinity revisited,''
		Phys.\ Rev.\ Lett.\  {\bf 105}, 111103 (2010)
		[arXiv:0909.2617 [gr-qc]].
		
		
		\bibitem{Strominger:2017zoo}
		A.~Strominger,
		``Lectures on the Infrared Structure of Gravity and Gauge Theory,''
		[arXiv:1703.05448 [hep-th]].
	
	
		\bibitem{Regge:1974zd}
		T.~Regge and C.~Teitelboim,
		``Role of Surface Integrals in the Hamiltonian Formulation of General Relativity,''
		Annals Phys. \textbf{88}, 286 (1974)
		doi:10.1016/0003-4916(74)90404-7
	
	
		\bibitem{Henneaux:2018cst}
		M.~Henneaux and C.~Troessaert,
		``BMS Group at Spatial Infinity: the Hamiltonian (ADM) approach,''
		JHEP \textbf{03}, 147 (2018)
		doi:10.1007/JHEP03(2018)147
		[arXiv:1801.03718 [gr-qc]].
	
	
		\bibitem{Henneaux:2019yax}
		M.~Henneaux and C.~Troessaert,
		``The asymptotic structure of gravity at spatial infinity in four spacetime dimensions,''
		[arXiv:1904.04495 [hep-th]].
		
			\bibitem{Ananth:2020ngt}
		S.~Ananth, L.~Brink and S.~Majumdar,
		``BMS algebra as an extension of the Poincar\'e symmetry in light-cone gravity,''
		[arXiv:2012.07880 [hep-th]].
		
		\bibitem{Bengtsson:2012dw}
		A.~K.~H.~Bengtsson, L.~Brink and S.~S.~Kim,
		``Counterterms in Gravity in the Light-Front Formulation and a D=2 Conformal-like Symmetry in Gravity,''
		JHEP \textbf{03}, 118 (2013)
		doi:10.1007/JHEP03(2013)118
		[arXiv:1212.2776 [hep-th]].
		
		
		\bibitem{Ananth:2017xpj} 
		S.~Ananth, L.~Brink, S.~Majumdar, M.~Mali and N.~Shah,
		``Gravitation and quadratic forms,''
		JHEP {\bf 1703}, 169 (2017)
		[arXiv:1702.06261 [hep-th]].
		
		
	
			\bibitem{Scherk:1974zm}
		J.~Scherk and J.~H.~Schwarz,
		``Gravitation in the light-cone gauge,''
		Gen. Rel. Grav. \textbf{6}, 537-550 (1975)
		doi:10.1007/BF00761962


		
		\bibitem{BCL}{I. Bengtsson, M. Cederwall and O. Lindgren, ``Light Front Actions for Gravity and Higher spins'', {\it {G\"oteborg-83-55}} (1983).}
		
		\bibitem{Ananth:2006fh}
		S.~Ananth, L.~Brink, R.~Heise and H.~G.~Svendsen,
		``The N=8 Supergravity Hamiltonian as a Quadratic Form,''
		Nucl. Phys. B \textbf{753}, 195-210 (2006)
		doi:10.1016/j.nuclphysb.2006.07.014
		[arXiv:hep-th/0607019 [hep-th]].

	
		\bibitem{Ananth:2008ik}
		S.~Ananth,
		``The Quintic interaction vertex in light-cone gravity,''
		Phys. Lett. B \textbf{664}, 219-223 (2008)
		doi:10.1016/j.physletb.2008.05.035
		[arXiv:0803.1494 [hep-th]].
		
		
		\bibitem{Bengtsson:1983pd}
		A.~K.~H.~Bengtsson, I.~Bengtsson and L.~Brink,
		``Cubic Interaction Terms for Arbitrary Spin,''
		Nucl. Phys. B \textbf{227}, 31-40 (1983)
		doi:10.1016/0550-3213(83)90140-2
		
				
		\bibitem{Dirac}
		P.~A.~M.~Dirac,
		``Forms of Relativistic Dynamics,''
		Rev. Mod. Phys. \textbf{21}, 392-399 (1949)
		doi:10.1103/RevModPhys.21.392
				
		\bibitem{Duval:2014uva}
		C.~Duval, G.~Gibbons and P.~Horvathy,
		``Conformal Carroll groups and BMS symmetry,''
		Class. Quant. Grav. \textbf{31} (2014), 092001
		[arXiv:1402.5894 [gr-qc]].
		
		\bibitem{Deser:1977hu}
		S.~Deser and C.~Teitelboim,
		``Supergravity Has Positive Energy,''
		Phys. Rev. Lett. \textbf{39}, 249 (1977)
		doi:10.1103/PhysRevLett.39.249
		
		\bibitem{Witten:1981mf}
		E.~Witten,
		``A Simple Proof of the Positive Energy Theorem,''
		Commun. Math. Phys. \textbf{80}, 381 (1981)
		doi:10.1007/BF01208277

		
		\bibitem{Ashtekar:1986yd}
		A.~Ashtekar,
		``New Variables for Classical and Quantum Gravity,''
		Phys. Rev. Lett. \textbf{57}, 2244-2247 (1986)
		doi:10.1103/PhysRevLett.57.2244

		
		\bibitem{SAMHV1}
		S. Ananth, S. Kovacs and S. Parikh,
		``A manifestly MHV Lagrangian for $N=4$ Yang-Mills,"
		JHEP \textbf{05}, 051 (2011) 
		[arXiv:1101.3540 [hep-th]].

		\bibitem{SAMHV2}
		S. Ananth,
		``Spinor helicity structures in higher spin theories,"
		JHEP \textbf{11}, 089 (2012) 
		[arXiv:1101.3540 [hep-th]].

		
		\bibitem{Strominger:2014pwa}
		A.~Strominger and A.~Zhiboedov,
		``Gravitational Memory, BMS Supertranslations and Soft Theorems,''
		JHEP \textbf{01}, 086 (2016)
		doi:10.1007/JHEP01(2016)086
		[arXiv:1411.5745 [hep-th]].

		

		\bibitem{Donnay:2018neh}
		L.~Donnay, A.~Puhm and A.~Strominger,
		``Conformally Soft Photons and Gravitons,''
		JHEP \textbf{01}, 184 (2019)
		doi:10.1007/JHEP01(2019)184
		[arXiv:1810.05219 [hep-th]].


	\bibitem{ABM1}
		S. Ananth, L. Brink and S. Majumdar,
		``$E_8$ in $N=8$ supergravity in four dimensions,''
		JHEP \textbf{01}, 086 (2016)
		doi:10.1007/JHEP01(2016)086
		[arXiv:1411.5745 [hep-th]].

		

		\bibitem{ABM2}
		S. Ananth, L. Brink and S. Majumdar,
		``Exceptional versus superPoincar\'e algebra as the defining symmetry of maximal supergravity,''
		JHEP \textbf{01}, 024 (2018)
		[arXiv:1711.09110 [hep-th]].
		
		
			
		\bibitem{Ananth:2015tsa}
		S.~Ananth, L.~Brink and M.~Mali,
		``Yang-Mills theories and quadratic forms,''
		JHEP \textbf{08}, 153 (2015)
		doi:10.1007/JHEP08(2015)153
		[arXiv:1507.01068 [hep-th]].
		
		
		\bibitem{Ananth:2020mws}
		S.~Ananth, C.~Pandey and S.~Pant,
		``Higher spins, quadratic forms and amplitudes,''
		JHEP \textbf{07}, no.07, 100 (2020)
		doi:10.1007/JHEP07(2020)100
		[arXiv:2005.10376 [hep-th]].
		
		
		\bibitem{Awada:1985by}
		M.~A.~Awada, G.~W.~Gibbons and W.~T.~Shaw,
		``Conformal Supergravity, Twistors and the super-BMS group'',
		Annals Phys. \textbf{171}, 52 (1986)
		doi:10.1016/S0003-4916(86)80023-9.

	\end{thebibliography}
\end{document}